\documentclass[review]{elsarticle}
\usepackage{lineno,hyperref}
\usepackage{amsmath,graphicx,makecell}
\usepackage{enumitem}
\usepackage{threeparttable}
\usepackage{booktabs}

\usepackage{algorithmic}
\usepackage[linesnumbered, ruled, vlined]{algorithm2e}
\SetAlFnt{\small}
\SetEndCharOfAlgoLine{}
\usepackage{xcolor}

\SetCommentSty{myComment}

\newcommand{\tabincell}[2]{\begin{tabular}{@{}#1@{}}#2\end{tabular}}

\usepackage[margin=2.5cm]{geometry}
\modulolinenumbers[5]
\journal{Applied Energy}
\usepackage{setspace}

\begin{document}

\begin{frontmatter}
\title{\textbf{Quantitative Assessment of U.S. Bulk Power Systems and Market Operations during the COVID-19 Pandemic}}
\author{
	Guangchun~Ruan,
	Jiahan~Wu,
	Haiwang~Zhong,
	Qing~Xia,
	Le~Xie
}
\begin{abstract}
	Starting in early 2020, the novel coronavirus disease (COVID-19) severely affected the U.S., causing substantial changes in the operations of bulk power systems and electricity markets. In this paper, we develop a data-driven analysis to substantiate the pandemic's impacts from the perspectives of power system security, electric power generation, electric power demand and electricity prices. Our results suggest that both electric power demand and electricity prices have discernibly dropped during the COVID-19 pandemic. Geographical variances in the impact are observed and quantified, and the bulk power market and power system operations in the northeast region are most severely affected. All the data sources, assessment criteria, and analysis codes reported in this paper are  available on a GitHub repository.
	
\end{abstract}
\end{frontmatter}

\section{Introduction} \label{SEC-INTRO} 
\subsection{Background and Motivation}
The novel coronavirus disease (COVID-19) outbreak was declared a global pandemic by the World Health Organization on March 11, 2020~\cite{RN14}. Two days later, the U.S. federal government issued a national emergency proclamation~\cite{RN15}. A series of emergency measures, however, failed to stop the rapid spread of COVID-19, and the U.S. soon became a new epicenter of the global outbreak~\cite{RN17}. To slow the spread of COVID-19, all states have implemented various policy interventions~\cite{RN28}, e.g., lockdown orders, social distancing measures, which directly caused an unprecedented reduction of commercial and industrial electricity consumption. While society is still trying to adapt to the changes brought by COVID-19, it is becoming evident that this public health crisis has had a far greater impact than originally anticipated by most experts~\cite{RN42,RN48,RN50}. 

This paper pays special attention to the pandemic's impact on U.S. bulk power systems and wholesale electricity markets. According to an overview~\cite{RN32} by the Federal Energy Regulatory Commission (FERC), there are seven regional transmission organizations, or wholesale electricity markets, in the U.S.---California (CAISO), Midcontinent (MISO), New England (ISO-NE), New York (NYISO), Pennsylvania-New Jersey-Maryland Interconnection (PJM), Southwest Power Pool (SPP), and Texas (ERCOT).

\begin{figure*}[t]
	\centering
	\includegraphics[width=0.95\textwidth]{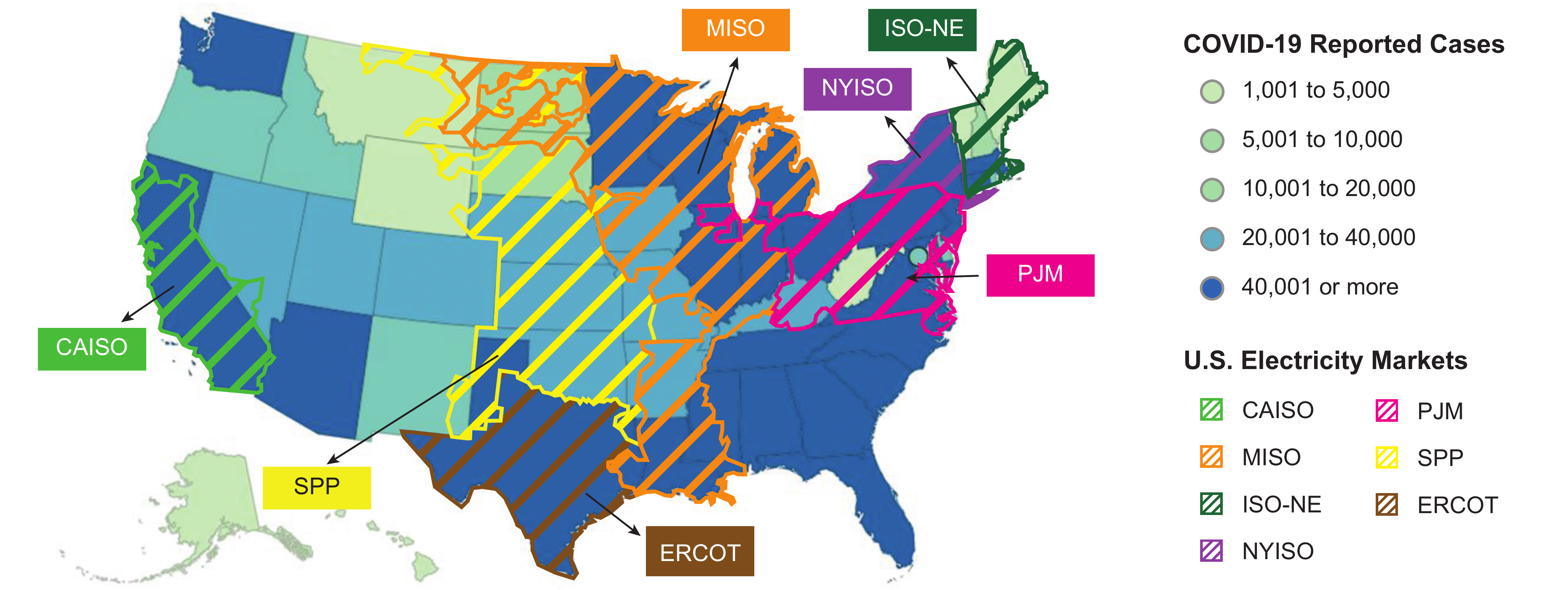}
	\caption{Confirmed COVID-19 cases as of July 20, 2020, in the U.S. and electricity marketplaces. The seven electricity markets cover the hardest-hit states, including New York, New Jersey, Illinois, California, Massachusetts, and Texas. Data source: U.S. Centers for Disease Control and Prevention (CDC) and Federal Energy Regulatory Commission (FERC).}
	\label{fig-covid-map}
\end{figure*}

Owing to their high quality and timely release, electricity market data are ideal for tracking the potential impacts of COVID-19. The existing marketplaces in the U.S. also largely cover most of the hotspot states. Fig.~\ref{fig-covid-map} shows confirmed COVID-19 cases as of July 20, 2020, at both the state level (heat map zones)~\cite{RN13} and the electricity marketplace level (shaded zones)~\cite{RN32}. 
% Quality control editor: Please ensure that the intended meaning has been maintained in the edits of the previous sentence and in similar instances.
The seven hardest-hit states and many dark blue areas~(over $40$ thousand confirmed cases) are properly covered by the existing U.S. electricity markets. These observations motivate us to make full use of the available market data to investigate how the electricity markets and power systems are affected.

\subsection{Related Literature and Analysis}
In late March, several reports discussed some pandemic-related changes. Reference~\cite{RN25, RN24} provided early observations of different electricity markets, and reference~\cite{RN22} commented on the additional issues of reliability risk, reduced bill payments, and delayed investment activities.

The Electric Power Research Institute (EPRI) reported more comprehensive observations from around the world~\cite{RN16}. The demand changes observed in Europe, the U.S., and China were classified based on restriction severity, and full details about New York state were given as well. The IEEE Power \& Energy Society released a report~\cite{RN45} collecting worldwide experiences and practices to mitigate the pandemic's adverse effects. In particular, this report presented much evidence regarding  electricity consumption, peak demand, and the generation mix. Reference~\cite{RN49} investigated the socioeconomic and technical problems faced by utility companies under different global scenarios, and the Indian power system was scrutinized as a case study.

The Energy Information Administration (EIA) released a short-term energy outlook~\cite{RN27} in the first quarter and pointed out the uncertain impact on electricity generation. This report also found continuously low wholesale prices and increasing residential electricity consumption in the U.S. Pecan Street company~\cite{RN23} made further efforts to monitor $113$ homes in Austin, Texas. Residential demand was high all day long, making the ``duck curve'' smoother than it has been over the past few years. Refrigerators appeared to work overtime, but electric vehicles were taking a long rest. Reference~\cite{RN46} analyzed day-ahead load forecasting under strict social distancing restrictions, and the mobility data were quite helpful in boosting prediction accuracy.

COVID-19 impact reports, mainly on electricity consumption, are periodically released by independent system operators (ISOs). In~\cite{RN61}, ERCOT applied a backcast model to estimate the load reduction and found weekly drops in energy use from $4 \sim 5\%$ in late April and $3 \sim 4\%$ in mid-May to approximately $1\%$ after mid-June. Electricity consumption in Texas was also lower during the early morning hours. A similar methodology was implemented by ISO-NE~\cite{RN62}, and they also found that the demand impact after mid-June was very limited. PJM reported that the peak impact in July was noticeably easing, which was possibly due to gradual reopening as well as increasing weather sensitivity~\cite{RN63}. Additionally, the NYISO forecasting team stated that commercial load reduction was a leading driver of low electricity demand~\cite{RN64}. Approximately $16\%$ of transmission outages planned by MISO were moved in the past few months, mainly because of the COVID-19 pandemic~\cite{RN65}. A CAISO report showed that power grid reliability was not affected by the stay-at-home order~\cite{RN66}. 

Reference~\cite{RN19} stated that shortages of photovoltaic modules, labor, and government purchases might heavily postpone solar installation. Reference~\cite{RN47} claimed that  countries rich in renewable energy were expected to see rapid increases in their shares of clean energy. In~\cite{RN52}, the authors estimated a $9.20\%$ reduction in global electricity production in 2020, and they calculated the change in global emissions accordingly. To navigate the crisis, reference~\cite{RN44} established a policy framework, including the short-term immediate response, mid-term economic recovery, and long-term energy transition. Similarly, reference~\cite{RN41} warned that COVID-19 could have a deep and negative impact on long-term innovation in clean energy if policy responses failed to take effect. Furthermore, COVID-19 assistance to target the energy insecurity of the low-income population was recommended ~\cite{RN40}.

In summary, COVID-19-related studies are still in their infancy for the following reasons: 1) limited depth and scope of study. For example, many power system--related indices have not been investigated in great detail. 2) Lack of cross-regional comparison. Given the broad geographical impact of COVID-19, a comprehensive cross-regional study of U.S. bulk power systems and markets would offer many valuable insights for policy makers. 3) Difficulty of scientific reproduction. Limitations around open-access data and well-organized tools make the aforementioned study results difficult to reproduce or analyze. 

This paper has bridged a knowledge gap to fully understand how COVID-19 affects power systems and markets. Building upon our previous work~\cite{RN71}, we have substantially extended the results by further considering the impacts on power system security, energy supply, and electricity prices. A series of novel criteria are developed in this paper along with extensive quantitative details, visualization results, and discussions. In addition, all the data and code used in this paper are available in a ready-to-use format and are publicly shared to facilitate reproduction.

\begin{figure}[t]
	\centering
	\includegraphics[width=0.6\textwidth]{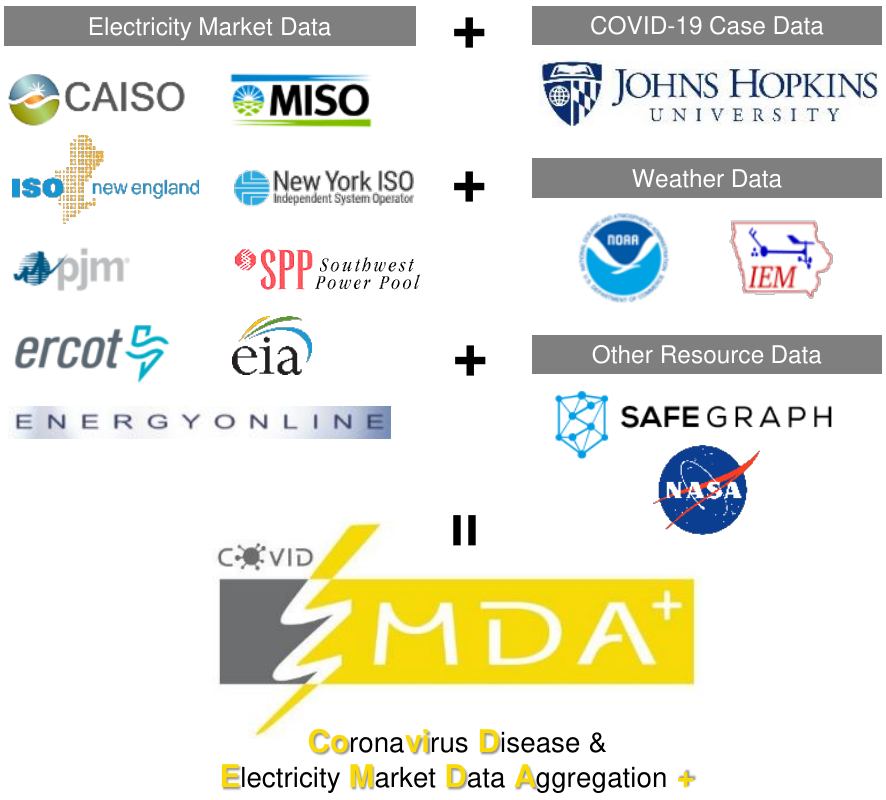}
	\caption{Data source description of the COVID-EMDA$^+$ data hub. This cross-domain data hub is designed to track the pandemic's impact on U.S. electricity markets and contributes to deepening a cross-domain understanding of the impacts of COVID-19. All analysis results reported in this paper are based on this data hub, which is open access and updated daily on GitHub~\cite{RN1}.}
	\label{fig-data-hub}
\end{figure}

\subsection{Open-Access Data Source and Code}
An open-access data hub, COVID-EMDA$^+$ (Coronavirus Disease and Electricity Market Data Aggregation)~\cite{RN1}, was established to track the potential impacts of COVID-19 on U.S. electricity markets and power systems. This data hub, with released data and parser tools, is updated daily to capture the latest situation. Data imports with web links avoid repetitive data refreshing and management processes. Note that all the data, parser tools, and analysis code are available on a GitHub repository~\cite{RN1,RN2}. 

To see more details, this data hub collects and harmonizes raw data from all existing U.S. electricity markets and combines them with other cross-domain data sources, e.g., public health data. Fig.~\ref{fig-data-hub} lists all the data sources that are integrated in the COVID-EMDA$^+$ data hub. The support team ensures data quality, handling most outliers and missing data by reviewing  backup data sources or historical trends. Several dedicated strategies are designed to consider different data features.

\subsection{Contributions and Paper Structure}
We summarize the key contributions of this paper as follows.
\begin{enumerate}
	\item This paper comprehensively assesses the impact of COVID-19 on the existing U.S. bulk power systems and markets. We conduct a data-driven analysis from the perspectives of power system security, electric power generation, electric power demand, and electricity prices. Many innovative criteria, e.g., excess change rates for renewables and abnormal price indices, are first developed in this paper.
	
	\item We find significant impacts on electric power demand and electricity prices in most regions. The impact is validated to be relatively diversified, both in intensity and dynamics, while the northeast region is found to be more sensitive than other places. Additionally, some evidence suggests that renewable generators are suffering extra curtailments due to COVID-19.
	
	\item All the data sources and code used in this paper are publicly shared. With daily updates and rigorous quality control, these ready-to-use resources help support various pandemic-related studies.
\end{enumerate}

The remainder of this paper is organized as follows. Section~\ref{SEC-SECURE} -- Section~\ref{SEC-PRC} are focused on the pandemic's impacts on power system security, electric power generation, electric power demand, and prices. Extensive results from different perspectives depict a full picture of the changes induced by COVID-19. Section~\ref{SEC-CONCL} concludes this paper.

\section{Impact on Power System Security} \label{SEC-SECURE}
An electricity market is an economic system that balances energy generation and demand in a power system. The first priority of market operators is monitoring the secure operation of the electric grid. This section explains how COVID-19 has affected  three aspects of power system operations, i.e., forecasting accuracy, congestion, and forced outages.

\subsection{Forecasting Accuracy}
The pandemic has caused significant changes in electricity consumption patterns, directly reducing the accuracy of demand forecasting. A poor prediction may increase operation risks and use many more flexibility resources. 

Here, we analyze day-ahead hourly demand forecasting by calculating the monthly mean absolute percentage error, which is defined as follows.
\begin{align}
	\label{eqn-forecast}
	e_{ym} = \frac{1}{N_m T} \sum_d \sum_t \left| \frac{\hat{D}_{ymdt} - D_{ymdt}}{D_{ymdt}} \right|, \quad \forall y,m
\end{align}
where $e_{ym}$ is the error for month~$m$ of year~$y$, $D_{ymdt}$ is the metered demand for year~$y$, month~$m$, day~$d$, and hour~$t$, and $\hat{D}_{ymdt}$ is the day-ahead hourly forecast. In addition, $N_m$ and $T$ are the number of days and hours, respectively.

Using the above formula, Table~\ref{tab-forecasting} compares the forecasting errors in 2020 and 2019. There is a clear trend in that the prediction errors are shrinking in most markets---larger errors occur before May, with accuracy improvements observed soon after. An interesting finding is that the forecast appears more accurate in July 2020 than in July 2019, implying a diminishing impact of COVID-19.
% Editor: Please ensure that the intended meaning has been maintained in the edits of the previous sentence.

Although the pandemic has posed new challenges for demand forecasting, the overall impact has been rather limited---even under the influence of COVID-19, the prediction errors are still within a tolerable range. Additionally, gradual improvement indicates that pandemic-induced risks have been properly managed with adequate data and knowledge accumulation.

\begin{table}[t] 
	\centering 
	\caption{Demand Forecasting Error in U.S. Electricity Markets [\%].} 
	\label{tab-forecasting} 
	\setlength\tabcolsep{12pt}  % 控制列宽 
	\begin{threeparttable} 
		\begin{tabular}{lccccc}
			\toprule
			Market & March  & April  & May  & June  & July  \\
			\midrule
			CAISO  & $3.4$ ($\bf{2.7}$) & $3.9$ ($\bf{2.8}$) & $6.0$ ($\bf{2.7}$) & $4.3$ ($\bf{4.1}$) & $3.9$ ($\bf{3.1}$) \\
			MISO   & $2.9$ ($\bf{1.6}$) & $3.0$ ($\bf{1.3}$) & $1.7$ ($\bf{1.3}$) & $2.4$ ($\bf{1.8}$) & $1.7$ ($\bf{1.6}$) \\
			ISO-NE & $2.5$ ($\bf{2.3}$) & $2.7$ ($\bf{2.5}$) & $3.1$ ($\bf{2.4}$) & $2.5$ ($\bf{2.4}$) & $\bf{2.1}$ ($3.1$) \\
			NYISO  & $\bf{2.3}$ ($2.8$) & $\bf{2.7}$ ($3.1$) & $\bf{2.0}$ ($3.2$) & $\bf{2.4}$ ($3.1$) & $\bf{2.0}$ ($2.8$) \\
			PJM    & $2.9$ ($\bf{1.9}$) & $2.8$ ($\bf{2.3}$) & $2.4$ ($\bf{1.7}$) & $2.7$ ($\bf{2.0}$) & $\bf{1.8}$ ($2.4$) \\
			SPP    & $4.9$ ($\bf{4.0}$) & $4.5$ ($\bf{3.8}$) & $3.9$ ($\bf{3.1}$) & $3.1$ ($\bf{3.0}$) & $4.2$ ($\bf{3.0}$) \\
			ERCOT  & $\bf{1.8}$ ($2.7$) & $2.3$ ($\bf{2.2}$) & $2.9$ ($\bf{2.3}$) & $\bf{2.5}$ ($3.0$) & $\bf{1.4}$ ($2.1$) \\
			\midrule
			Mean   & $3.0$ ($\bf{2.6}$) & $3.1$ ($\bf{2.6}$) & $3.1$ ($\bf{2.4}$) & $2.8$ ($2.8$) & $\bf{2.4}$ ($2.6$) \\
			\bottomrule
		\end{tabular}
		\begin{tablenotes}
			\item Note: The above data are forecasting errors in 2020 (outside parentheses) and 2019 (within parentheses). The smaller error items are highlighted in bold. We cover the results from March 1 to July 15 for both years.
		\end{tablenotes}
	\end{threeparttable}	 
\end{table}

\begin{figure}[t]
	\centering
	\includegraphics[width=0.5\textwidth]{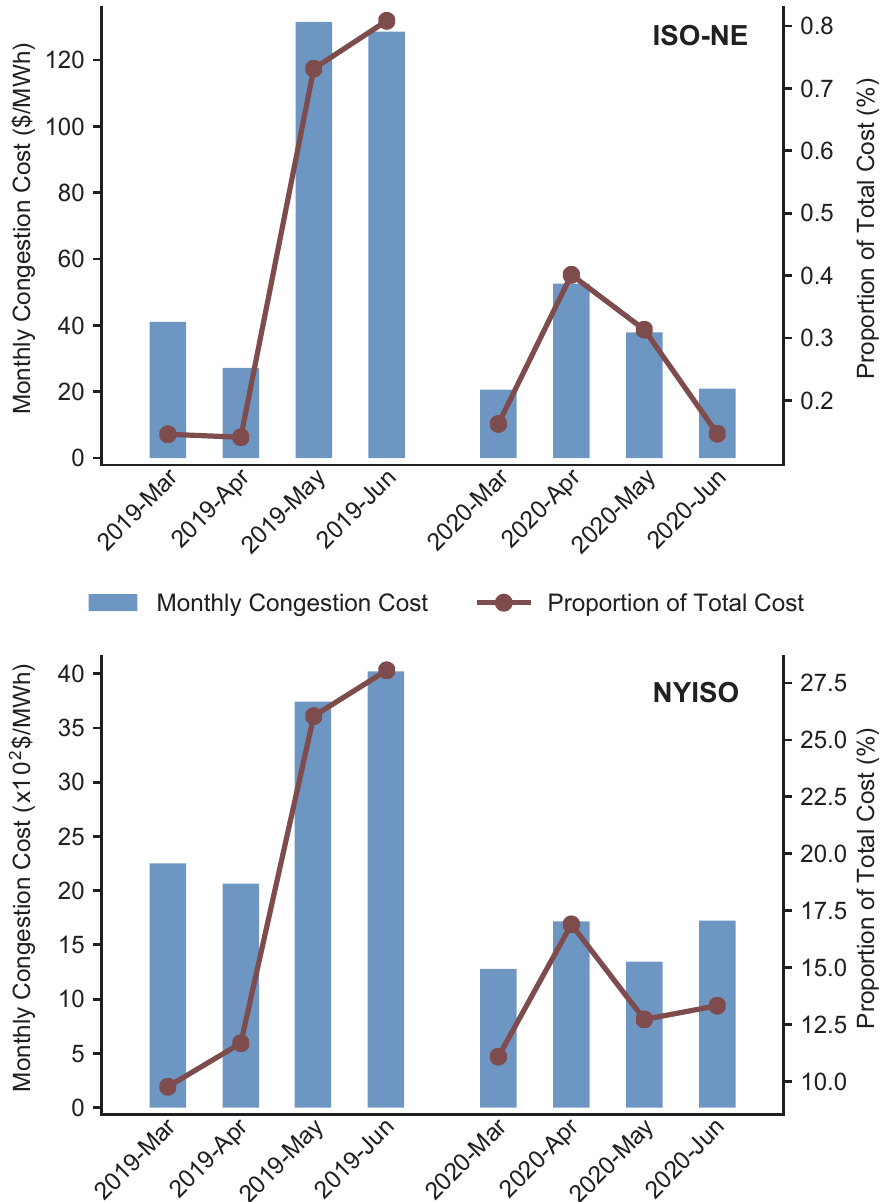}
	\caption{Monthly congestion statistics in ISO-NE and NYISO. This figure shows both the congestion cost (bar chart) and the associated proportion of total electricity cost (line chart). The results from March to June for 2019 and 2020 are provided for comparison, and a significant drop can be observed in May and June 2020.}
	\label{fig-congestion}
\end{figure}

\subsection{Congestion}
Transmission network congestion occurs when the available network capacity cannot satisfy the electric power demand without exceeding  safety requirements. A heavily congested network appears to be less reliable because the transmission capacity is often insufficient. Given that congestion is related to many factors, it might be effective to conduct an overall assessment by analyzing congestion price data.

Electricity market operators calculate locational marginal prices every day, and congestion prices are one important component. Market clearing theory states that these congestion prices are the shadow prices of the corresponding operational constraints. Based on this fact, congestion status can therefore be studied by analyzing the  congestion price data. 

This paper mainly focuses on monthly statistics and calculates the monthly congestion cost as follows.
\begin{align}
	\label{eqn-congestion}
	C_{ym}^\text{Cong} = \sum_d \sum_t \Big( \frac{1}{A} \sum_a \left| \lambda_{aymdt}^\text{Cong} \right| \Big), \quad \forall y,m
\end{align}
where $C_{ym}^\text{Cong}$ is the congestion cost per megawatt hour for year~$y$ and month~$m$.  $\lambda_{aymdt}^\text{Cong}$ is the congestion price in area~$a$ (representing a region in an electricity marketplace), and $A$ is the total number of areas. We apply an absolute function here because the congestion price may become negative when the associated power flow direction reverses.

Fig.~\ref{fig-congestion} shows monthly congestion statistics for  ISO-NE and NYISO. When compared with 2019 data,  congestion costs roughly decreased after the COVID-19 outbreak, especially in May and June. This improvement can be explained by the fact that electricity demand during the COVID-19 pandemic was lower than that during normal times (see Section~\ref{SEC-DMD} for more details)---this change is beneficial to transmission capacity (larger redundancy), and congestion is thus less likely to happen. For the proportion of total cost, one can find a slight increase in March and April followed by a significant drop in the next two months. Note that the total electricity prices have rapidly decreased during the COVID-19 pandemic (a more detailed discussion is available in Section~\ref{SEC-PRC}).

\begin{figure}[t]
	\centering
	\includegraphics[width=0.5\textwidth]{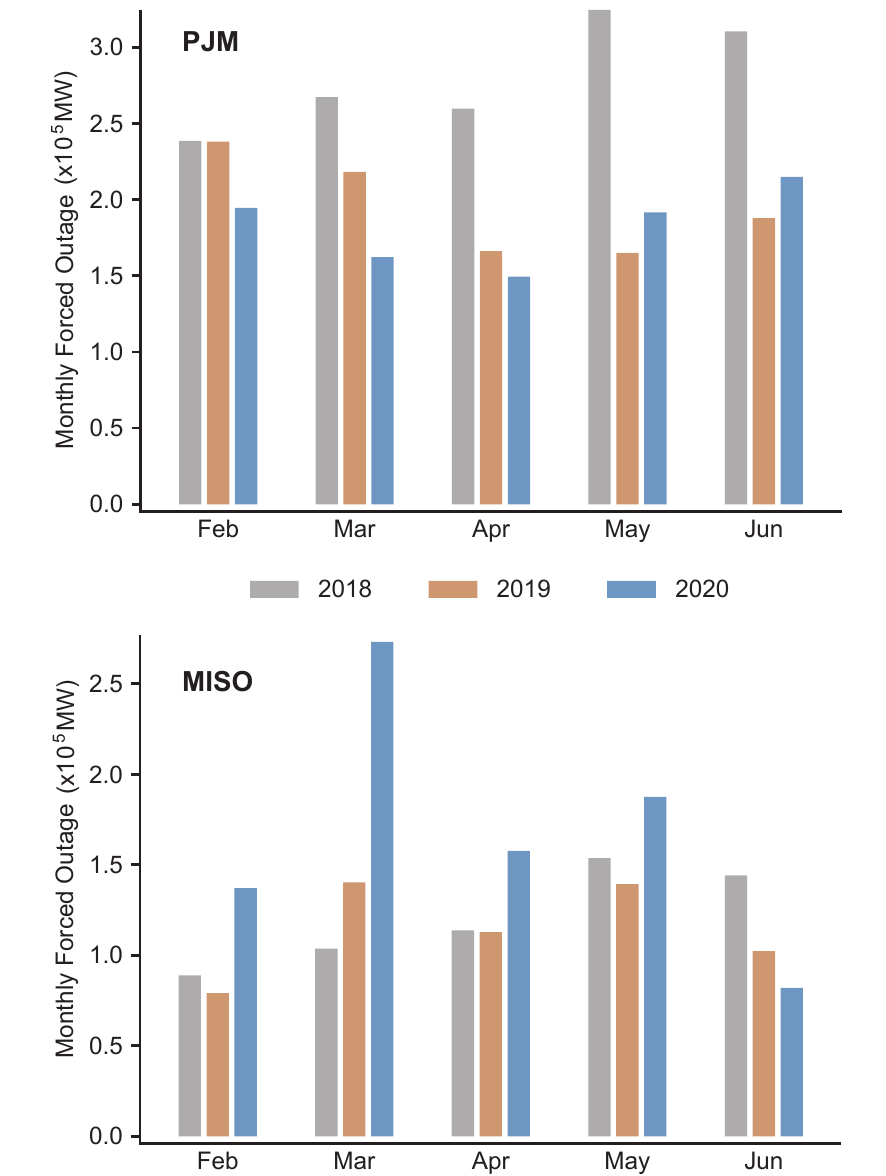}
	\caption{Monthly forced outage requirements in PJM and MISO. The data from February to June 2020 and the past two years are provided for comparison. Larger changes are found in MISO, especially in February and March.}
	\label{fig-outage}
\end{figure}

\subsection{Forced Outage}
A forced outage refers to an unexpected shutdown that occurs when some generators are unavailable to produce electricity. More forced outages generally produce higher risk and deteriorate power system reliability. From a technical perspective, we sum up the forced outage data for each month to develop a high-level criterion.

Fig.~\ref{fig-outage} shows the monthly forced outage results in PJM and MISO (only available in these two markets). In 2020, the total megawatt-hours of forced outages in PJM is similar to or slightly better than that of the previous two years. This is not true for MISO, especially in March, but an obvious improvement can be found later in June. A possible explanation for the abnormal observation in March may be the early hurricane season reported by meteorology analysis~\cite{RN57,RN58}. One can learn from the above results that, currently, no strong evidence supports a significant and consistent impact. In other words, COVID-19 might have more limited impacts than expected.

\section{Impact on Electric Power Generation} \label{SEC-GEN} 
This section analyzes changes in electric power generation during the COVID-19 pandemic with a special focus on structural changes as well as on renewable generation status.

\begin{figure}[t]
	\centering
	\includegraphics[width=0.6\textwidth]{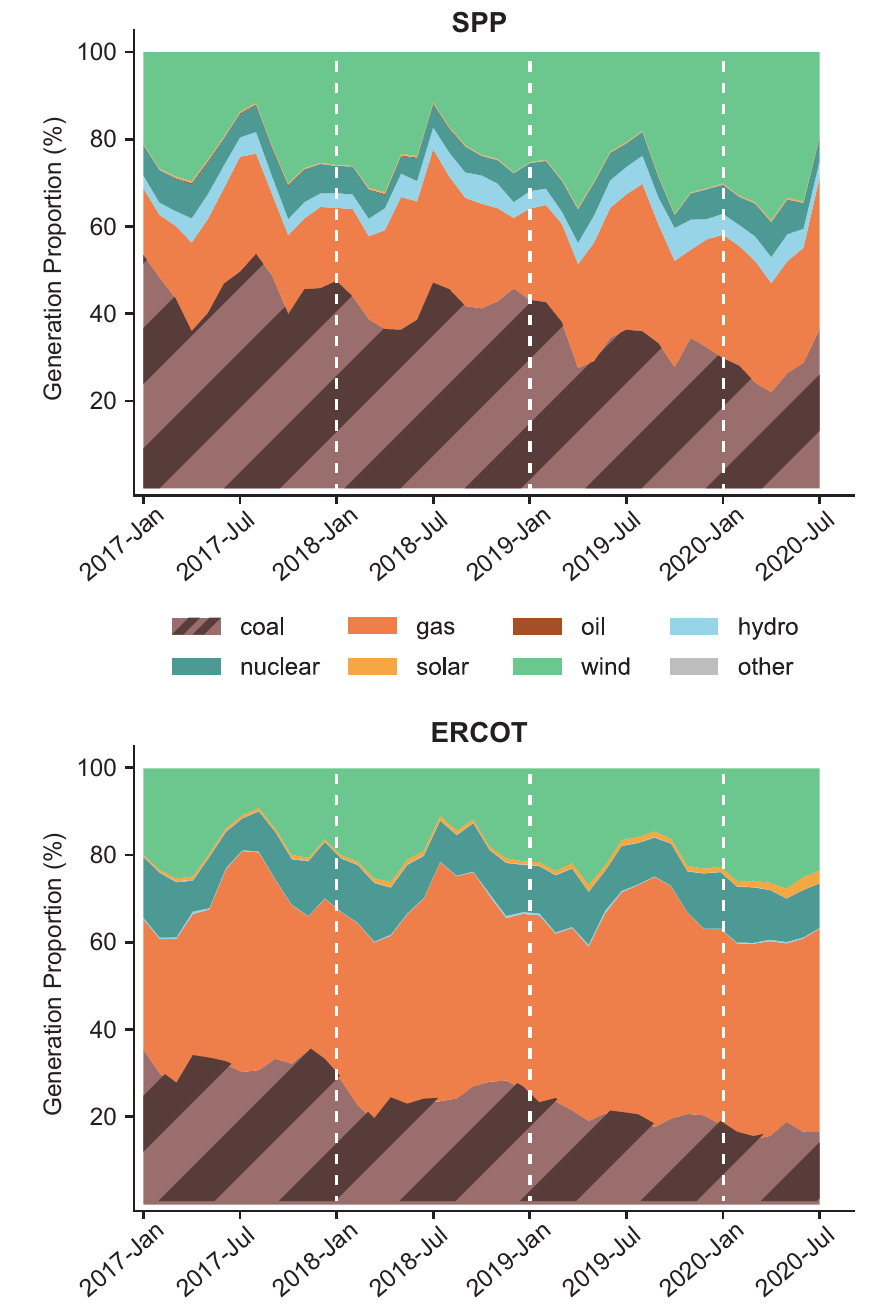}
	\caption{Electricity generation mix in SPP and ERCOT. The data are recorded from 2017 to 2020 for both markets, showing a clear trend of gradual replacement of carbon-intensive fuel (coal) by other cleaner fuel.}
	\label{fig-gen-mix}
\end{figure}

\subsection{Generation Mix}
The generation mix, or generation structure, refers to the combination and proportion of various types of generators. A key question is whether the generation mix is different during the COVID-19 pandemic than during the pre-pandemic era, and the answer will provide important evidence of a disproportional impact on different generators.

Fig.~\ref{fig-gen-mix} shows the generation mix results in SPP and ERCOT, and each figure is divided by dotted lines to show the observations for different years. In SPP, one can clearly observe that carbon-free generation (wind) and less carbon-intensive fuel (natural gas) have a slowly increasing market share, while coal-fired electricity generation continues dropping. A similar trend is also found in ERCOT, with natural gas becoming the major fuel source. We further test other markets and validate that this is an evolving trend rather than a coincidence, mainly because of ongoing clean energy plans. 

A further finding is that no significant difference in structural change (consistently a linear trend) is observed in all electricity markets during the COVID-19 pandemic. From a theoretical perspective, the generation mix is mainly determined by cost competition among different energy fuels. Although the pandemic should have a mild impact on relative fuel costs, cost competition is becoming more intense due to the suppression of total power demand. Coal-fired generators have experienced a tough time due to their high cost and low efficiency, while renewable generators are expected to gain market share. These inferences, however, are not totally compatible with our observations, and the difference may be explained by some specific dispatch strategies that keep the generation structure stable for safety concerns.

\begin{table}[t] 
	\centering 
	\caption{Proportion of Renewable Generation in U.S. Electricity Markets [\%].} 
	\label{tab-renewable} 
	\setlength\tabcolsep{12pt}  % 控制列宽 
	\begin{threeparttable} 
		\begin{tabular}{lcccc}
			\toprule
			Market & 2017     & 2018     & 2019     & 2020     \\
			\midrule
			CAISO  & $21.0$            & $23.8$            & $25.5$            & $26.1$ \\
			MISO   & $\phantom{0}8.3$  & $\phantom{0}7.4$  & $\phantom{0}9.1$  & $12.3$ \\
			ISO-NE & $\phantom{0}3.1$  & $\phantom{0}3.4$  & $\phantom{0}3.6$  & $\phantom{0}4.8$  \\
			NYISO  & $\phantom{0}3.2$  & $\phantom{0}2.6$  & $\phantom{0}3.2$  & $\phantom{0}3.4$  \\
			PJM    & $\phantom{0}2.7$  & $\phantom{0}2.6$  & $\phantom{0}3.2$  & $\phantom{0}3.9$  \\
			SPP    & $22.6$            & $23.7$            & $27.1$            & $33.1$ \\
			ERCOT  & $18.6$            & $20.5$            & $21.3$            & $27.8$ \\
			\midrule
			Mean   & $11.4$            & $12.0$            & $13.3$            & $15.9$ \\
			\bottomrule
		\end{tabular}
	\end{threeparttable}	 
\end{table} 

\begin{figure}[t]
	\centering
	\includegraphics[width=0.5\textwidth]{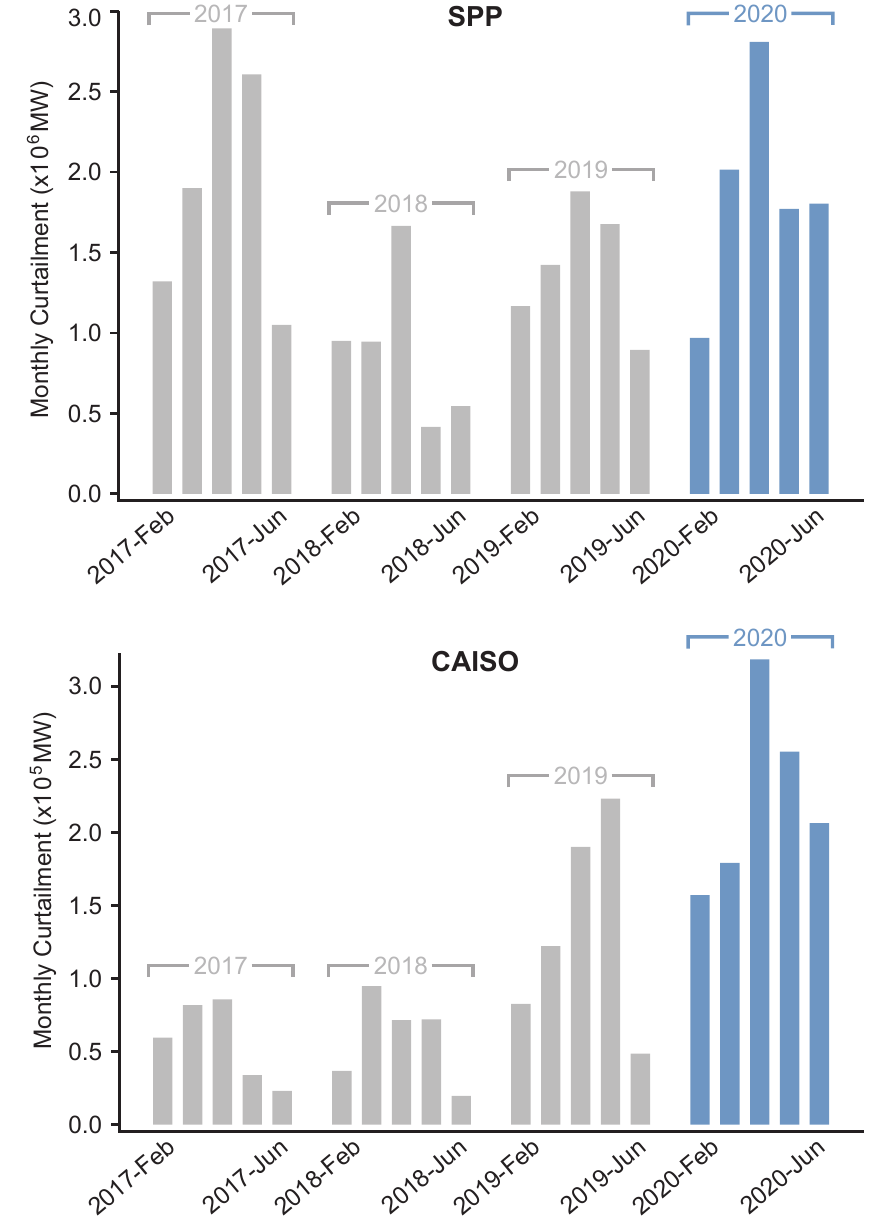}
	\caption{Monthly renewable curtailment in SPP and CAISO. This figure provides four-year data for comparison. When compared with those of the previous year, renewable energy curtailments are observed more frequently in 2020 for both markets. Additionally, renewable generators in CAISO have experienced a more severe impact than those in SPP.}
	\label{fig-curtailment}
\end{figure}

\subsection{Market Share and Curtailment of Renewable Generation}
Here, we continue the discussion of the previous subsection and provide more quantitative analysis to check whether renewable generators have obtained extra benefits (larger market share due to lower marginal cost) after the COVID-19 outbreak.

Table~\ref{tab-renewable} shows the annual market share of renewable energy from 2017 to 2020. One can trace the proportion change in a row and make the cross-market comparison in a column. Mild changes can be found in all electricity markets except ISO-NE and ERCOT. The market shares in CAISO and NYISO remain roughly unchanged, which may indicate a larger curtailment rate. 

We formulate an excess change rate to quantify the market share changes during the COVID-19 pandemic. The detailed formulation is given below.
\begin{align}
	\label{eqn-excess-rate}
	\eta = \Big( \frac{r_\text{2020}}{2 r_\text{2019} - r_\text{2018}} - 1 \Big) \times 100\%
\end{align}
where $\eta$ is the excess change rate, and $r_\text{2018} \sim r_\text{2020}$ are the market shares of renewable energy observed in 2018, 2019, and 2020. Assuming a linear growth rate, the denominator term~$(2 r_\text{2019} - r_\text{2018})$ represents the estimated market share without the pandemic. Formula~(\ref{eqn-excess-rate}) finally calculates a relative change rate between the observed market share and this estimation.

The excess change rates for all markets are as follows: CAISO~($-4.0\%$), MISO~($13.9\%$), ISO-NE~($26.3\%$), NYISO~($-10.5\%$), PJM~($2.6\%$), SPP~($8.5\%$) and ERCOT~($25.8\%$). These results match the above observations that the renewable generators in CAISO and NYISO might have lower market shares than expected. This outcome is probably due to some operational safety concerns; for example, an online report~\cite{RN53} found that solar generation during daylight hours in CAISO already outpaces the decreasing electricity demand caused by COVID-19.

 Further, we analyze renewable curtailment status in SPP and CAISO (only available in these two markets). Fig.~\ref{fig-curtailment} shows monthly renewable curtailment from 2017 to 2020, and one can observe increasing curtailment in 2020 for both markets when compared with that in 2019. Furthermore, CAISO tends to apply a more aggressive curtailment strategy than SPP, and this finding also matches the results of the excess change rates. Although market competition theory suggests that renewable generators can obtain extra benefits relative to their competitors, this situation may become quite complex and unclear when considering the curtailment issue.

From a long-term development perspective, renewable generators still face the barrier of high capital costs (high installing expenses), and they rely heavily on a stable and continuous cash flow.  This inherent vulnerability might cause more financial difficulties during the COVID-19 pandemic because federal subsidies might be reduced or postponed. More attention, in this respect, should be paid to the long-term financial influences on renewable energy.

\begin{figure*}[t]
	\centering
	\includegraphics[width=1\textwidth]{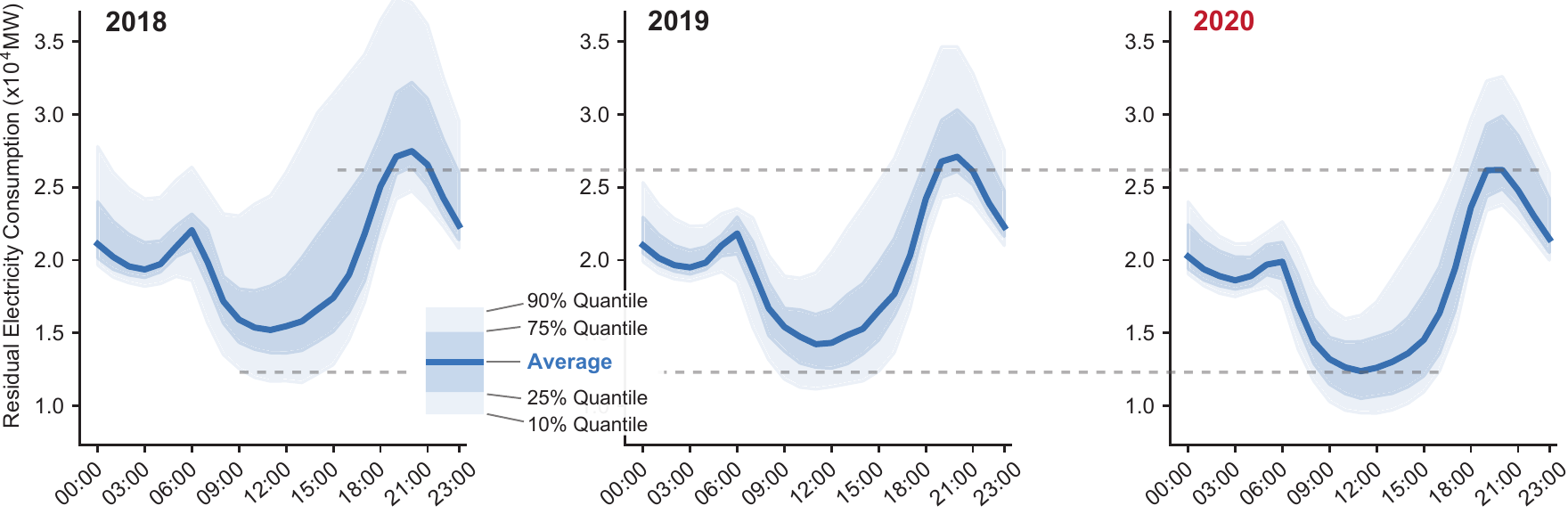}
	\caption{Duck curve changes in CAISO. Two-year historical data are provided with the $10\% \sim 90\%$ and $25\% \sim 75\%$ quantiles. A larger peak-valley difference and peak-valley ratio in 2020 can be verified when compared with the past two years.}
	\label{fig-duck-curve}
\end{figure*}

\subsection{Duck Curves and Ramping Requirements} \label{SUBSEC-DUCK-CURVE}
The duck curve was first developed in 2012 from a CAISO report, and it soon attracted wide interest in both academia and industry. This curve shows the daily imbalance between peak demand and renewable energy generation, and it can be expressed by the residual demand as follows.
\begin{align}
	\label{eqn-duck}
	R_{ymdt} = D_{ymdt} - G_{ymdt}^\text{Solar}, \quad \forall y,m,d,t 
\end{align}
where $R_{ymdt}$ is the residual electricity consumption (duck curve) for year~$y$, month~$m$, day~$d$, and hour~$t$, and $G_{ymdt}^\text{Solar}$ is the solar generation at the same time.

Note that the duck curve phenomenon (residual demand drop significantly at noon) merely happens in California because of the high penetration of solar generation. In other markets, e.g., ERCOT and SPP, solar grows fast but still accounts for a small proportion.

Fig.~\ref{fig-duck-curve} compares the duck curves from 2018 to 2020. The data from March to mid-July of each year are used. The average duck curve profiles are shown in the middle with two uncertain intervals~($25\% \sim 75\%$ and $10\% \sim 90\%$ quantile intervals). These uncertain intervals are calculated as follows: given a typical year~$y$ and hour~$t$, the marginal distribution of the residual demand profile data is expressed as $R_{ymdt} \sim \hat{F}_{yt}, \ \forall m, d$. Here, $\hat{F}_{yt}$ can be visualized in Fig.~\ref{fig-duck-curve} with four quantiles, i.e., $\hat{F}_{yt}^{-1} (10\%)$, $\hat{F}_{yt}^{-1} (25\%)$, $\hat{F}_{yt}^{-1} (75\%)$ and $\hat{F}_{yt}^{-1} (90\%)$.

Fig.~\ref{fig-duck-curve} clearly exhibits a shift down in the duck curve of 2020, mainly because of the reduction in electricity consumption. The valley part shrinks slightly more than the peak part, resulting in an enlarged peak-valley difference and a larger peak-valley ratio. This pattern implies an increase in the ramping requirement and operation risk as well. We then consider the maximal hourly ramp-up and ramp-down requirements and calculate the average values for each year. The average ramp-up requirement in 2020 is $4284.5$~megawatts, higher than 2019 ($3886.2$~megawatts) and 2018 ($3497.7$~megawatts). For ramping down, we again find the highest requirement, $3444.2$~megawatts, in 2020, compared with 2019 ($3212.5$~megawatts) and 2018 ($3195.9$~megawatts).

\begin{figure*}[t]
	\centering
	\includegraphics[width=0.9\textwidth]{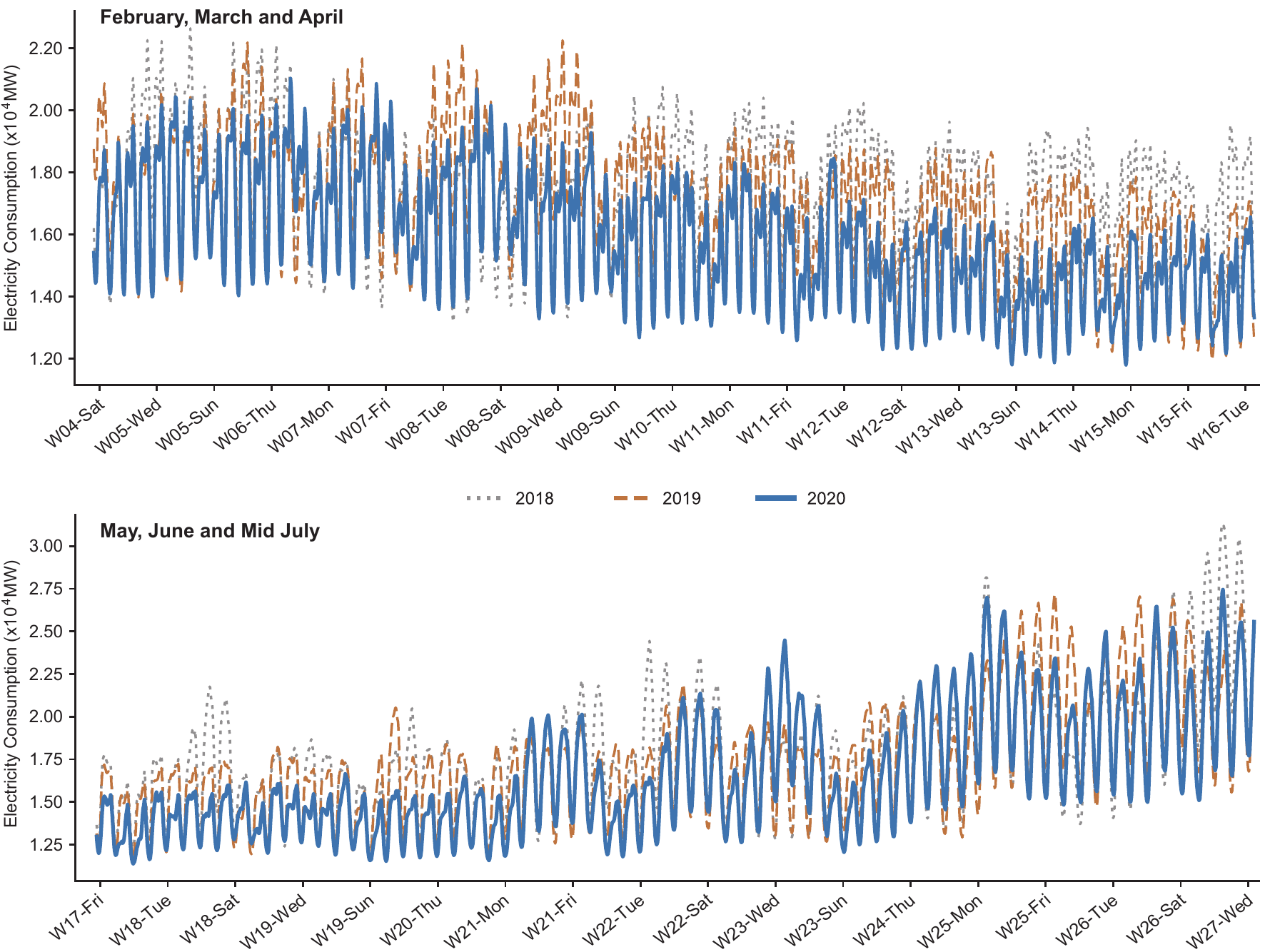}
	\caption{Hourly electricity consumption in New York state from 2018 to 2020. The data from February to May are aligned by weekday. Compared with the past two years, a significant drop can be observed since March 2020, and this trend continued until a rebound in the last week of May.}
	\label{fig-3yr-load}
\end{figure*}

\section{Impact on Electric Power Demand} \label{SEC-DMD} 
This section concentrates on the changes in electricity consumption across all U.S. electricity markets. A typical case study of New York state and a cross-market comparison are fully discussed.

\subsection{Demand Profiles in New York State}
The demand side is believed to have experienced a significant change after the COVID-19 outbreak, and we first focus on the epicenter: NYISO in New York state. 

Fig.~\ref{fig-3yr-load} shows electricity consumption from February to mid-July in the last three years. These three-year data are properly aligned to allow comparisons between the same weekdays. Technically, the week alignment is implemented by using the following formula.
\begin{align}
	\label{eqn-3yr-load}
	\begin{bmatrix} m' \\ d' \end{bmatrix} = 
	W_{y-1}^{-1} \bigg( W_y \Big( \begin{bmatrix} m \\ d \end{bmatrix} \Big) \bigg), \quad \forall y,m,d
\end{align}
where $y,m,d$ are the associated year, month, and day, respectively. $W_y (\cdot)$ is a transform function that converts a calendar date to a week-weekday format, while $W_{y-1} (\cdot)$ can do an opposite transform for the previous year. 

Formula~(\ref{eqn-3yr-load}) finally derives an aligned date pair $(y, m, d)$ and $(y-1, m', d')$, and both dates are the same weekday of the same week in the corresponding year. This alignment  helps eliminate the impact of week patterns.

As shown in Fig.~\ref{fig-3yr-load}, electricity consumption starts dropping in March and remains low until late May. During this period, the average hourly reduction rate is $6.4\%$, and the maximal rate is $25.3\%$ when taking the 2019 data as the baseline. A rebound phenomenon can be clearly observed in  the last week of May (week 21 in Fig.~\ref{fig-3yr-load}), mainly because of gradual reopening policies. This pattern continued until July, by which the demand had recovered to almost normal levels.

\begin{figure*}[p]
	\centering
	\includegraphics[width=1\textwidth]{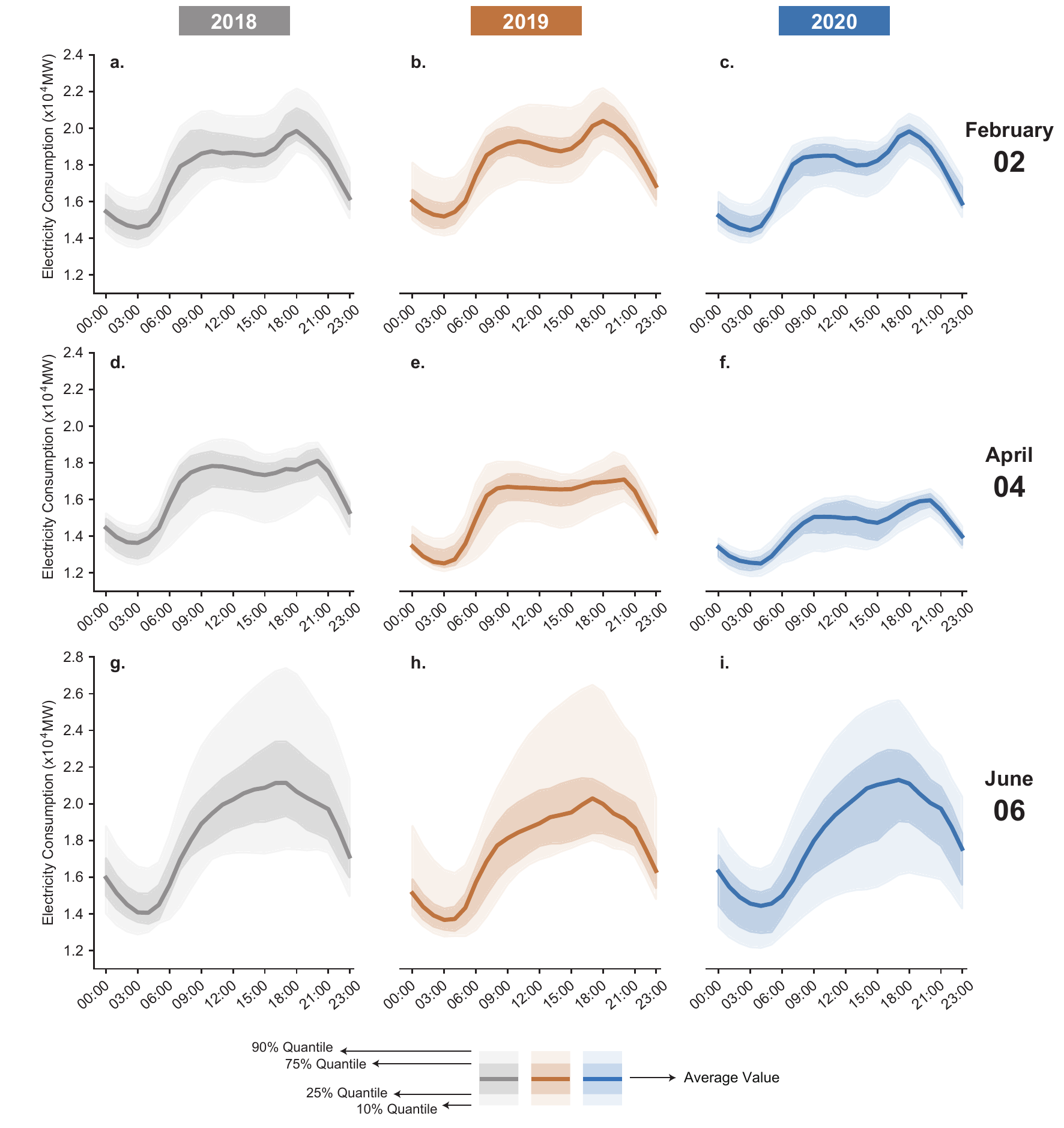}
	\caption{Comparison of daily averaged electricity demand profiles in NYISO. Each column shows the profile for one year (2018--2020), and each row shows the profile for one month (February--May). The $10\% \sim 90\%$, $25\% \sim 75\%$ quantiles are also given. The results indicate that a significant change can be found in March 2020.}
	\label{fig-3yr-load-profile}
\end{figure*}

\begin{figure*}[t]
	\centering
	\includegraphics[width=1\textwidth]{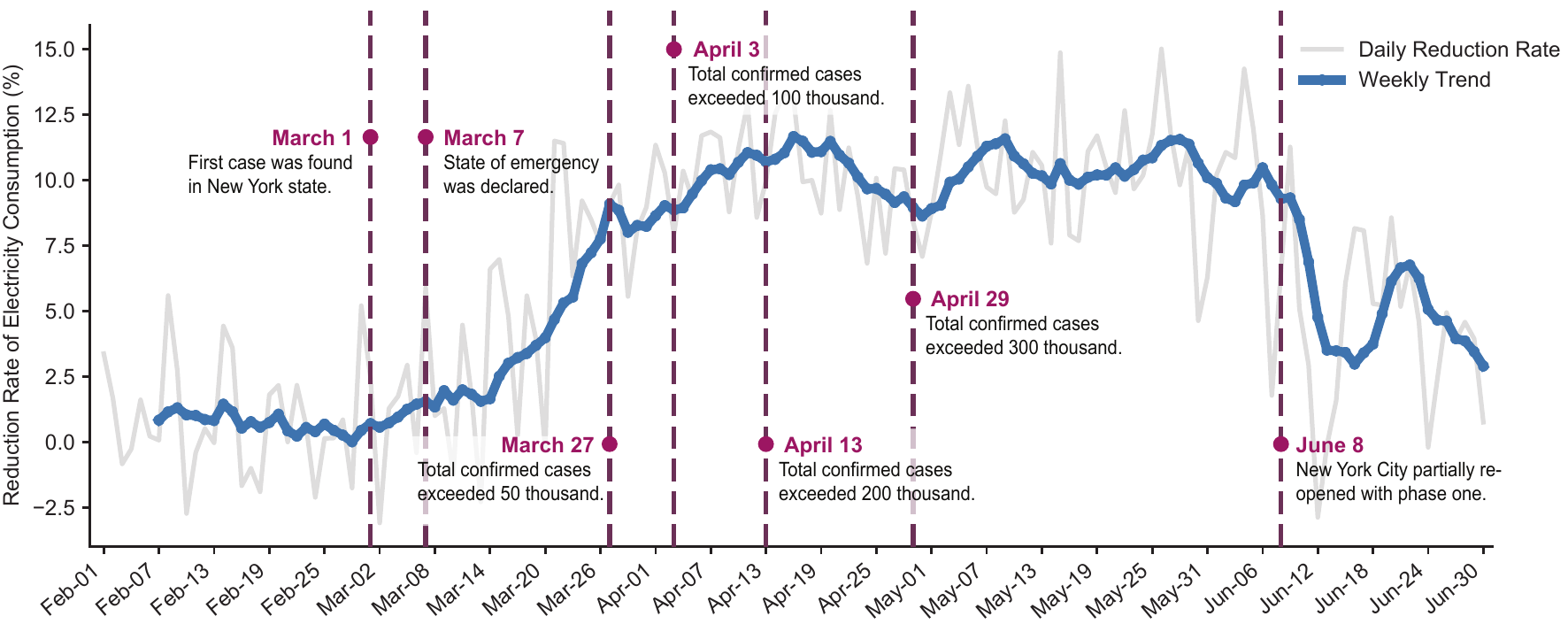}
	\caption{Time-varying electricity demand reduction rate in NYISO during the COVID-19 pandemic. The weekly trend is calculated by means of a moving average technique. Some important events for New York state are highlighted in the timeline.}
	\label{fig-load-timeline}
\end{figure*}

We next plot and compare the demand profiles in each month of each year. For all the subfigures in Fig.~\ref{fig-3yr-load-profile}, the average daily load profiles are plotted in the middle with two uncertain intervals. These intervals are calculated by means of a method similar to that in Subsection~\ref{SUBSEC-DUCK-CURVE}.

Fig.~\ref{fig-3yr-load-profile} compares February, April, and June. One can find a significant drop from February to April followed by a rapid rebound in June. Further comparisons are made by calculating the coefficient of variance for all averaged profile curves. The results show that the curves in April 2020 are flatter than in the previous two years, with a $14.5\%$ drop in the coefficient of variance. One can also observe a relatively smooth morning ramping in April 2020, which is a rather beneficial feature for system operation. The daily load profiles seem to be less stochastic in April 2020, while the profiles in the other two months exhibit more fluctuation. Moreover, very few curve differences can be identified in June 2020 when compared with 2018 and 2019. 

To further eliminate the impact of weather factors, we apply an ensemble backcast model to our previous work~\cite{RN71} to produce an adjusted estimation of the demand reduction rate. Fig.~\ref{fig-load-timeline} shows the estimations with several date labels highlighting some important events during the COVID-19 pandemic. We find a good match between demand changes and those important COVID-19 events in New York state, especially in the fast-developing period from late February to mid-April. This observation, from another perspective, reveals a hidden relationship between electricity consumption data and public health data.

\subsection{Demand Change Across Different Markets}
For cross-market comparison, we extend the previous analysis to other markets and highlight some key observations as follows.

The results show that the northeast region experienced the greatest influence, while a limited impact can be observed in the southern area. This finding is probably due to different electricity consumption behaviors, and electricity use in the southern area shifted rather than decreased. Before June, large load changes often occurred in the hardest-hit areas in terms of confirmed cases. An exception is  late June, when an abnormal growth in case numbers happened in Texas. In terms of the dynamic trend, CAISO is the first market to start recovering in mid-May, nearly one or two weeks earlier than other markets. We also find an obvious rebound in the load reduction rate in CAISO (which first drops from late April to early May and rebounds for a few days before a second drop after mid-May) and ERCOT (which first drops from mid-May to late May and rebounds until early June before soon dropping again).

\begin{figure}[t]
	\centering
	\includegraphics[width=0.45\textwidth]{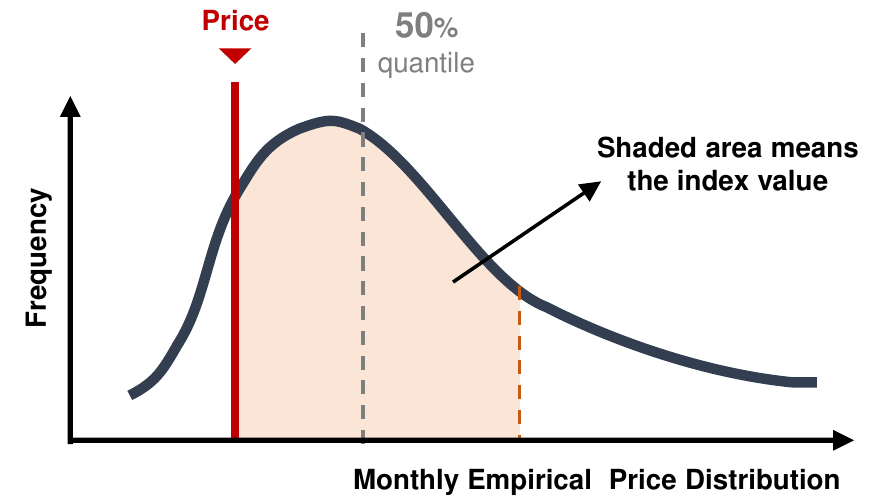}
	\caption{Statistical illustration of the proposed abnormal price index. The basic idea is to show how near a price observation is to the mean value. This index can reliably eliminate some stochastic factors when assessing price changes. The index value lies in $[0, 1]$, and a larger value means a higher possibility of abnormality.}
	\label{fig-index}
\end{figure}

\section{Impact on Wholesale Electricity Price} \label{SEC-PRC} 
This section analyzes the pandemic's impact on wholesale electricity prices, specifically, day-ahead locational marginal prices. We first develop an abnormal price index and then conduct a full assessment of all U.S. electricity markets.

\subsection{Abnormal Price Index}
Locational marginal prices play a crucial role in electricity markets and contribute to balancing generation and demand efficiently. Since these prices are relatively stochastic in nature, it is unsatisfactory to analyze them in the same way as electricity demand (the uncertainty intervals will be very wide). Thus, the abnormal price index is developed to provide a more reliable tool for price analysis. This index is defined as follows.
\begin{align}
	\label{eqn-index}
	I ( \lambda_{ymdt} ) = \big| 2 \hat{F}_m ( \lambda_{ymdt} ) - 1 \big|, \quad \forall y,m,d,t 
\end{align}
where $\lambda_{ymdt}$ is the day-ahead locational marginal price of year~$y$, month~$m$, day~$d$, and hour~$t$. $I (\cdot)$ is the proposed index, which lies between zero and one. This index quantifies the abnormality of a typical price, and a larger index value represents a more unusual observation. $\hat{F}_m (\cdot)$ is the cumulative distribution function for the prices in month~$m$. We will show in the next subsection that monthly distributions are more stable for analyzing these price changes.

The statistical meaning of the above index is explained by the probability density function shown in Fig.~\ref{fig-index}. For a given price, the index denotes a possibility that measures how close this price is to the mean value. For example, if a price observation is located within the $25\%$ and $75\%$ quantiles, we obtain an index value below $0.5$, which is considered normal. 

Since this abnormal price index has a clear statistical meaning and a concise expression, it is extremely suitable to analyzing price changes with uncertainty.

\begin{figure}[t]
	\centering
	\includegraphics[width=0.5\textwidth]{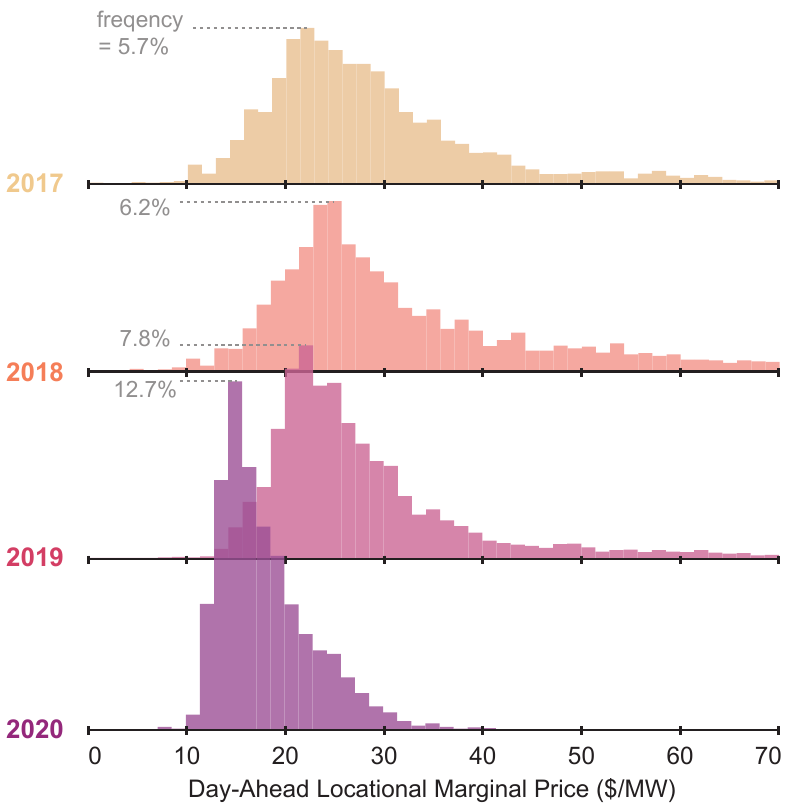}
	\caption{Day-ahead locational marginal price distribution in ISO-NE. The data from March to mid-July 2020 and the past four years are analyzed. Obvious changes in price distribution can be verified by a lower mean value as well as a higher peak (peak frequencies are provided).}
	\label{fig-lmp-distrib}
\end{figure}

\subsection{Price Distribution in the New England Area}
As a typical example, Fig.~\ref{fig-lmp-distrib} shows the price distributions for several years in the New England area. We select the price data from March to mid-July to derive the frequency graph and highlight the highest frequencies. A clear observation is that the distributions for 2017 to 2019 look similar but quite different from that for 2020 (this remains true for a single month). A more concentrated distribution is found in 2020 with a shrinking mean value, which reflects continuous observations of low prices. To see this, the average prices from March to mid-July drop from approximately US$\$30$ before 2020 to US$\$18.13$ in 2020.

Extensive testing beyond Fig.~\ref{fig-lmp-distrib} is conducted, and we finally find that the monthly price distributions are stable enough to eliminate the stochastic influences but remain highly  precise. The daily price data are more sensitive to some unexpected outliers than electricity demand, and those outliers are unfortunately common. For example, in April 2018, the average price at 12 PM is US$\$42.87$, $66\%$ higher than that in 2019, and the uncertain interval between the $25\%$ and $75\%$ quantiles is US$\$31.12$, $72.6\%$ of the average price (but for electricity demand, this number never exceeds $21\%$).

\begin{figure*}[t]
	\centering
	\includegraphics[width=1\textwidth]{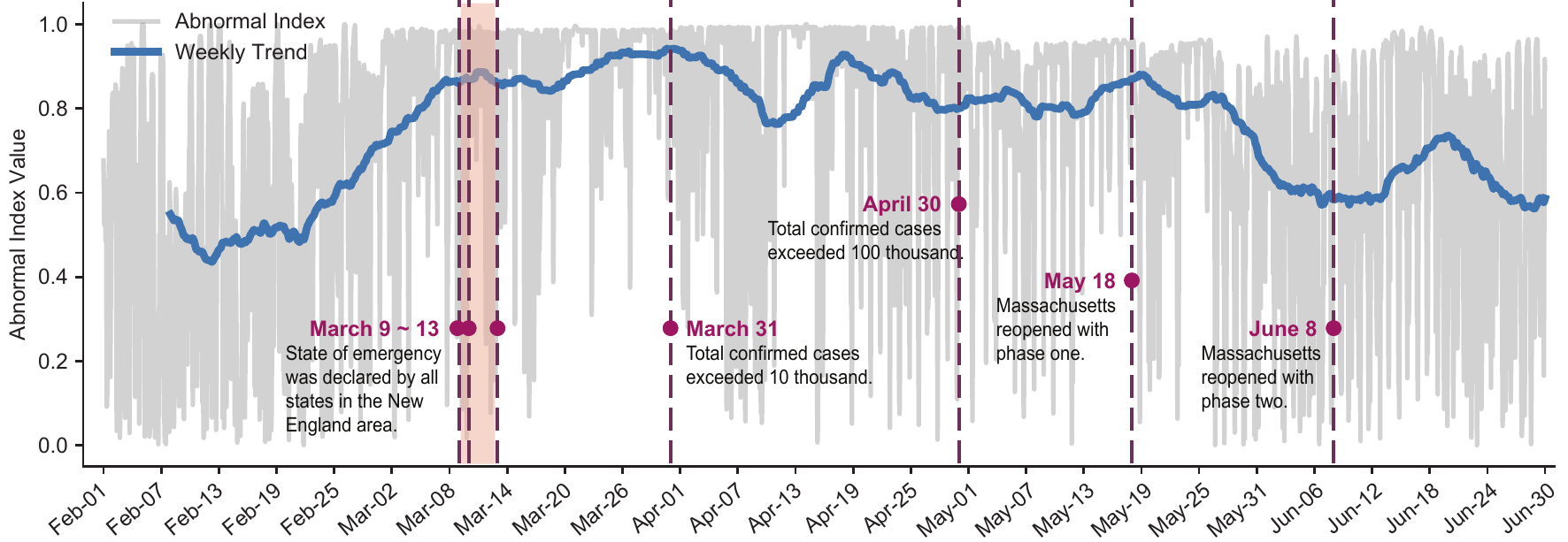}
	\caption{Time-varying values of the abnormal price index in ISO-NE during the COVID-19 pandemic. The weekly trend is obtained by means of a moving average technique. The declaration of a state of emergency order and other important events are highlighted in the timeline. The increasing abnormal price index values before March imply that prices were already going down prior to the pandemic.}
	\label{fig-lmp-timeline}
\end{figure*}

This finding also implies that the proposed abnormal price index can show reliable performance by analyzing the price distributions. Fig.~\ref{fig-lmp-timeline} plots the trend of the abnormal price index from February to mid-July 2020. Some important events related to COVID-19 are highlighted, including  declaration dates for state of emergency orders.

A surprising finding is that the index values are already very high before the state of emergency declarations, which indicates that another factor is driving the price down before the COVID-19 outbreak. The most likely factor is the natural gas price collapse, which can be verified by comparing Fig.~\ref{fig-lmp-timeline} with the gas price curve~\cite{RN54}. We next substantiate that the electricity prices are experiencing a double impact of the gas price and COVID-19 pandemic. To this end, we calculate the Pearson correlation coefficient between the abnormal 	index and gas price, before and after the declaration dates for state of emergency. The results show that the coefficient drops from $-0.502$ (moderate negative relationship) to $-0.045$ (weak relationship), and this is probably due to an offset effect of the above two factors. An illustrative evidence in late April is that the abnormal index still remains high when the gas price quickly rebounds (it is expected that COVID-19 has an opposite effect).

\subsection{Price Distribution Change Across Different Areas}
To conduct a cross-market comparison, we apply a Wasserstein probability distance metric to quantify the price change during the COVID-19 pandemic. The price change for year~$y$ can be formulated as follows.
\begin{align}
	\label{eqn-price-dist}
	s_y = \text{WD} \Big( \text{Vec}(\lambda_{ymdt}) - \text{Vec}(\lambda_\text{hist}) \Big), \quad \forall y
\end{align}
where $s_y$ denotes the price change for year~$y$ that is calculated by a Wasserstein distance function $\text{WD}(\cdot)$. $\text{Vec}(\cdot)$ is a vectorization function to place all the price data from March to mid-July in a one-dimensional array. $\lambda_\text{hist}$ represents all the historical price data accordingly.

Table~\ref{tab-lmp} presents the price change results together with some public health statistics. Significant price distribution changes can be found in the hardest hit areas, especially for NYISO, ISO-NE, and PJM. Additionally, those markets in the northeast region experienced a larger price drop than the other areas. Since gas prices across the nation are similar, the findings in Table~\ref{tab-lmp} roughly capture the intensity of the pandemic's impact across different marketplaces.

\begin{table}[t] 
	\centering 
	\caption{Price Changes and Public Health Statistics in U.S. Electricity Markets.} 
	\label{tab-lmp} 
	\setlength\tabcolsep{5pt}  % 控制列宽 
	\begin{threeparttable} 
		\begin{tabular}{lccc}
			\toprule
			Market 
			& \tabincell{c}{Price Distribution \\ Distance} 
			& \tabincell{c}{Total Confirmed \\ Cases [$\times 10^3$]}
			& \tabincell{c}{Virus Infection \\ Rate [\%]}  \\
			\midrule
			CAISO  & $\phantom{0}7.848$  & $355.3$  & $0.90$  \\
			MISO   & $\phantom{0}8.678$  & $771.1$  & $1.28$  \\
			ISO-NE & $11.774$            & $186.6$  & $1.26$  \\
			NYISO  & $10.407$            & $404.0$  & $1.54$  \\
			PJM    & $11.548$            & $702.3$  & $1.05$  \\
			SPP    & $\phantom{0}8.582$  & $117.9$  & $0.77$  \\
			ERCOT  & $\phantom{0}9.680$  & $305.5$  & $1.05$  \\
			\bottomrule
		\end{tabular}
	\end{threeparttable}	 
\end{table}

\section{Conclusion} \label{SEC-CONCL} 
This paper conducts a comprehensive assessment of the impact of the COVID-19 pandemic on all U.S. electricity markets and associated bulk power systems. Drawing on the COVID-EMDA$^+$ data hub, we provide some strong evidence that the power sector was highly influenced from March to May, entering a recovery period after June.  We also find very diverse impacts in different marketplaces, so market-specific analysis is critically important. Electric power demand and prices are more heavily affected than power grid operations and electric power generation.

Based on current observations, the impact of COVID-19 may not be a high risk for existing electricity markets, but we should pay more attention to possible shocks in the near future (e.g., from a second wave in late July) and some mid- to long-term influences. It may also be important to focus on disproportionate impacts on utility companies and consumers. All these issues require considerable effort from the whole society.

Although COVID-19 will not disappear immediately, the energy community can minimize potential adverse impacts by monitoring the situation. To this end, the open-access COVID-EMDA$^+$ data hub and analysis methods used in this paper can be relatively useful. Our future work involves developing novel methods to understand the complex economy-energy relationship during the COVID-19 pandemic, which can provide further insights to prepare for an uncertain future.

\section*{Reference}
\Urlmuskip=0mu plus 1mu\relax  % split url


\begin{thebibliography}{10}
	
\bibitem{RN14}
{World Health Organization}, ``Coronavirus disease {(COVID-19)} pandemic,''
Available:
\url{https://www.who.int/emergencies/diseases/novel-coronavirus-2019}
[Online], 2020.

\bibitem{RN15}
D.~J. Trump, ``Proclamation on declaring a national emergency concerning the
novel coronavirus disease {(COVID-19)} outbreak,'' Available:
\url{https://www.whitehouse.gov/presidential-actions/proclamation-declaring-national-emergency-concerning-novel-coronavirus-disease-covid-19-outbreak/}
[Online], 2020.

\bibitem{RN17}
E.~Dong, H.~Du, and L.~Gardner, ``An interactive web-based dashboard to track
{COVID-19} in real time,'' {\em The Lancet Infectious Diseases}, 2020.

\bibitem{RN28}
{National Association of Regulatory Utility Commissioners}, ``State response
tracker,'' Available:
\url{https://www.naruc.org/compilation-of-covid-19-news-resources/state-response-tracker/}
[Online], 2020.

\bibitem{RN42}
I.~M. Peters, C.~Brabec, T.~Buonassisi, J.~Hauch, and A.~M. Nobre, ``The impact
of {COVID-19} related measures on the solar resource in areas with high
levels of air pollution,'' {\em Joule}, 2020.

\bibitem{RN48}
A.~Mandal, R.~Roy, D.~Ghosh, S.~Dhaliwal, A.~Toor, S.~Mukhopadhyay, and
A.~Majumder, ``{COVID-19} pandemic: Sudden restoration in global
environmental quality and its impact on climate change,'' {\em EnerarXiv
	preprint}, 2020.

\bibitem{RN50}
C.~Magazzino and N.~Schneider, ``The relationship between air pollution and
{COVID-19-related} deaths: An application to three french cities,'' {\em
	EnerarXiv preprint}, 2020.

\bibitem{RN32}
{Federal Energy Regulatory Commission}, ``Electric power markets: national
overview,'' Available:
\url{https://www.ferc.gov/market-assessments/mkt-electric/overview.asp}
[Online], 2020.

\bibitem{RN13}
{Centers for Disease Control and Prevention}, ``Cases of coronavirus disease
{(COVID-19)} in the {U.S.},'' Available:
\url{https://www.cdc.gov/coronavirus/2019-ncov/cases-updates/cases-in-us.html}
[Online], 2020.

\bibitem{RN25}
R.~Walton, ``Utilities beginning to see the load impacts of {COVID-19} as
economic shutdown widens,'' Available:
\url{https://www.utilitydive.com/news/utilities-are-beginning-to-see-the-load-impacts-of-covid-19-as-economic-sh/574632/}
[Online], 2020.

\bibitem{RN24}
R.~Conrad, ``Electricity demand in the time of {COVID-19},'' Available:
\url{https://www.forbes.com/sites/greatspeculations/2020/03/30/electricity-demand-in-the-time-of-covid-19/#18e3fd007e86}
[Online], 2020.

\bibitem{RN22}
A.~J. Lawson, ``{COVID-19}: Potential impacts on the electric power sector,''
Available: \url{https://crsreports.congress.gov/product/pdf/IN/IN11300}
[Online], 2020.

\bibitem{RN16}
{Electric Power Research Institute}, ``Demand impacts and operational and
control center practices,'' Available:
\url{http://mydocs.epri.com/docs/public/covid19/3002018602R2.pdf} [Online],
2020.

\bibitem{RN45}
A.~Paaso, S.~Bahramirad, J.~Beerten, E.~Bernabeu, B.~Chiu, B.~Enayati,
B.~Hederman, L.~Jones, Y.~Jun, H.~Koch, and J.~C. Montero, ``Sharing
knowledge on electrical energy industry’s first response to {COVID-19},''
Available: \url{https://resourcecenter.ieee-pes.org/technical-publications/
	white-paper/PES_TP_COVID19_050120.html} [Online], 2020.

\bibitem{RN49}
V.~Senthilkumar, K.~Reddy, and U.~Subramaniam, ``{COVID-19}: Impact analysis
and recommendations for power and energy sector operation,'' {\em EnerarXiv
	preprint}, 2020.

\bibitem{RN27}
{Energy Information Administration}, ``Short-term energy outlook,'' Available:
\url{https://www.eia.gov/outlooks/steo/report/electricity.php} [Online],
2020.

\bibitem{RN23}
S.~Hinson, ``{COVID-19} is changing residential electricity demand,''
Available: \url{https://www.pecanstreet.org/2020/04/covid/} [Online], 2020.

\bibitem{RN46}
Y.~Chen, W.~Yang, and B.~Zhang, ``Using mobility for electrical load
forecasting during the {COVID-19} pandemic,'' {\em arXiv preprint arXiv:
	2006.08826}, 2020.

\bibitem{RN61}
C.~Opheim and S.~Parody, ``{COVID-19} load impact analysis ({ERCOT}),''
Available:
\url{http://www.ercot.com/content/wcm/lists/200201/ERCOT_COVID-19_Analysis_July_7.pdf}
[Online], 2020.

\bibitem{RN62}
{Independent System Operator of New England}, ``Estimated impacts of {COVID-19}
on {ISO New England} demand,'' Available:
\url{https://www.iso-ne.com/static-assets/documents/2020/07/isone-covid-19-update-07-07-2020.pdf}
[Online], 2020.

\bibitem{RN63}
A.~Gledhill, ``Recent load impacts ({PJM}),'' Available:
\url{https://www.pjm.com/~/media/committees-groups/committees/pc/2020/20200707/20200707-item-09-recent-load-impacts.ashx}
[Online], 2020.

\bibitem{RN64}
{New York Independent System Operator}, ``{COVID-19} related updates,''
Available: \url{https://www.nyiso.com/covid} [Online], 2020.

\bibitem{RN65}
{Midcontinent Independent System Operator}, ``{COVID-19} impact to load and
outage coordination,'' Available:
\url{https://cdn.misoenergy.org/20200705\%20COVID\%2019\%20Impacts\%20to\%20MISO\%20Load\%20and\%20Outage457884.pdf} [Online], 2020.
	
\bibitem{RN66}
{California Independent System Operator}, ``{COVID-19} impacts to {California
	ISO} load and markets: March 17–july 5, 2020,'' Available:
\url{http://www.caiso.com/Documents/COVID-19-Impacts-ISOLoadForecast-Presentation.pdf#search=covid\%20impact} [Online], 2020.
	
\bibitem{RN19}
K.~Das, ``Impact of {COVID-19} pandemic into solar energy generation sector,''
Available: \url{https://ssrn.com/abstract=3580341} [Online], 2020.

\bibitem{RN47}
D.~Chiaramonti and K.~Maniatis, ``Security of supply, strategic storage and
{COVID-19}: Which lessons learnt for renewable and recycled carbon fuels, and
their future role in decarbonizing transport?,'' {\em Applied Energy},
vol.~271, p.~115216, 2020.

\bibitem{RN52}
Y.~Yusup, J.~S. Kayode, M.~I. Ahmad, C.~S. Yin, M.~S. M.~N. Hisham, and H.~M.
Isa, ``Atmospheric {CO2} and total electricity production before and during
the nation-wide restriction of activities as a consequence of the {COVID-19}
pandemic,'' {\em arXiv preprint arXiv:2006.04407}, 2020.

\bibitem{RN44}
B.~Steffen, F.~Egli, M.~Pahle, and T.~S. Schmidt, ``Navigating the clean energy
transition in the {COVID-19} crisis,'' {\em Joule}, vol.~4, no.~6,
pp.~1137--1141, 2020.

\bibitem{RN41}
K.~T. Gillingham, C.~R. Knittel, J.~Li, M.~Ovaere, and M.~Reguant, ``The
short-run and long-run effects of {COVID-19} on energy and the environment,''
{\em Joule}, 2020.

\bibitem{RN40}
M.~Graff and S.~Carley, ``{COVID-19} assistance needs to target energy
insecurity,'' {\em Nature Energy}, vol.~5, no.~5, pp.~352--354, 2020.

\bibitem{RN71}
G.~Ruan, D.~Wu, X.~Zheng, S.~Sivaranjani, L.~Xie, H.~Zhong, and C.~Kang, ``A
cross-domain approach to analyzing the short-run impact of {COVID-19} on the
{U.S.} electricity sector,'' Available at SSRN:
\url{https://ssrn.com/abstract=3631498} [Online], 2020.

\bibitem{RN1}
G.~Ruan, X.~Zheng, D.~Wu, J.~Wu, S.~A. Alimi, and L.~Xie, ``{COVID-EMDA+} data
hub,'' Available:
\url{https://github.com/tamu-engineering-research/COVID-EMDA} [Online], 2020.

\bibitem{RN2}
G.~Ruan, ``Supplementary codes,'' Available:
\url{https://github.com/tamu-engineering-research/COVID-EMDA/tree/master/supplementary/impact-assessment-paper}
[Online], 2020.

\bibitem{RN57}
{National Oceanic and Atmospheric Administration}, ``Busy {Atlantic} hurricane
season predicted for 2020,'' Available:
\url{https://www.noaa.gov/media-release/busy-atlantic-hurricane-season-predicted-for-2020}
[Online], 2020.

\bibitem{RN58}
{Midcontinent Independent System Operator}, ``{MISO} expects adequate resources
for summer season,'' Available:
\url{https://www.misoenergy.org/about/media-center/miso-expects-adequate-resources-for-summer-season/}
[Online], 2020.

\bibitem{RN53}
J.~S. John, ``California renewables curtailments surge as coronavirus cuts
energy demand,'' Available:
\url{https://www.greentechmedia.com/articles/read/california-renewable-curtailments-spike-as-coronavirus-reduces-demand}
[Online], 2020.

\bibitem{RN54}
{BusinessInsider}, ``Natural gas price,'' Available:
\url{https://markets.businessinsider.com/commodities/natural-gas-price}
[Online], 2020.
	
\end{thebibliography}
\end{document}